\documentclass{article}
\usepackage{spconf,amsmath,graphicx}
\usepackage{bm, amssymb}
\usepackage{subfigure}
\usepackage{booktabs}
\usepackage{caption} 
\usepackage{cite}
\title{Speaker Embedding-aware Neural Diarization for Flexible Number of Speakers with Textual Information}

\name{Zhihao Du, Shiliang Zhang, Siqi Zheng, Weilong Huang, Ming Lei}
\address{Speech Lab, Alibaba Group, China\\
\texttt{\{neo.dzh,sly.zsl\}@alibaba-inc.com}}

\begin{document}
\ninept

\maketitle

\begin{abstract}
Overlapping speech diarization is always treated as a multi-label classification problem.
In this paper, we reformulate this task as a single-label prediction problem by encoding the multi-speaker labels with power set.
Specifically, we propose the speaker embedding-aware neural diarization (SEND) method, which predicts the power set encoded labels according to the similarities between speech features and given speaker embeddings.
Our method is further extended and integrated with downstream tasks by utilizing the textual information, which has not been well studied in previous literature.
The experimental results show that our method achieves lower diarization error rate than the target-speaker voice activity detection.
When textual information is involved, the diarization errors can be further reduced.
For the real meeting scenario, our method can achieve 34.11\% relative improvement compared with the Bayesian hidden Markov model based clustering algorithm.
\end{abstract}

\begin{keywords}
Speaker diarization, single-label predication, power-set encoding, textual information
\end{keywords}

\section{Introduction}
\label{sec:intro}
Speaker diarization aims at answering the question ``who spoken when''.
It is an important technique for a wide range of real-world applications, such as the speaker-attributed automatic speech recognition (SA-ASR) \cite{JaninBEEGMPPSSW03,CarlettaABFGHKKKKLLLMPRW05,BarkerWVT18}.

A typical approach for speaker diarization is based on the extraction and clustering of speaker embeddings.
In this approach, the audio recording is first split into several segments by voice activity detection (VAD). 
Then, speaker embeddings, such as i-vector \cite{DehakKDDO11}, d-vector \cite{WanWPL18} and x-vector \cite{SnyderGSPK18}, are extracted from the segments.
To obtain the diarization results, speaker embeddings are partitioned into clusters by unsupervised clustering algorithms, such as k-means \cite{DimitriadisF17}, spectral clustering \cite{NingLTH06} and agglomerative hierarchical clustering (AHC) \cite{Garcia-RomeroSS17}.
Recently, the variational Bayesian hidden Markov model is introduced to perform the clustering of x-vector sequences (VBx) \cite{DiezBWRC19} and achieves superior performance on CALLHOME \cite{CastaldoCDLV08}, AMI \cite{CarlettaABFGHKKKKLLLMPRW05} and DIHARD-II \cite{RyantCCCDGL19} datasets.

There are two drawbacks of clustering based approaches. First, they have trouble handling overlapping speech due to the speaker-homogeneous assumption of each segment.
Second, clustering algorithms do not aim at minimizing the diarization errors directly, which are performed in an unsupervised manner.
To overcome the drawbacks, end-to-end neural diarization (EEND) \cite{FujitaKHNW19} is introduced and improved by combining with the self-attention mechanism \cite{FujitaKHXNW19}, Conformer architecture \cite{Liu2021Conformer} and encoder-decoder attractor \cite{HoriguchiF0XN20}.
EEND-based models are always optimized with the permutation-invariant training (PIT) strategy \cite{KolbaekYTJ17}, which has trouble handling large number of speakers at the training stage.

Another approach is the target-speaker voice activity detection (TSVAD) \cite{MedennikovKPKKS20}, which is designed to minimize the diarization errors directly and free from the label permutation problem. 
However, TSVAD requires a fixed and known number of speakers for both training and test, which limits its application to real conversations \cite{he21c_interspeech}.
More importantly, both EEND and TSVAD treat the overlapping speech diarization as a multi-label classification problem.
In this view, the activities of different speakers are regarded as independent events, which can be not satisfied in real meeting scenarios.
Another drawback of this formulation is that their performance heavily depends on the manually selected activation thresholds.

In this paper, we attempt to solve the problems of label permutation and threshold selection by reformulating the overlapping speech diarization task as a single-label prediction problem.
Under this formulation, the speaker embedding-aware neural diarization (SEND) is proposed.
In SEND, we first calculate the similarities between speech features and given speaker embeddings.
Then, the similarities are fed to a neural network to model the relationship between speaker activities. 
Unlike EEND and TSVAD, the network is trained to predict a one-hot label, which represents the speaker combination for each frame.
The one-hot label is derived from the independent labels of each speaker by using the power set encoding.
In addition, SEND is further extended to utilize the textual information with a self-attention based encoder, which much benefits the downstream application SA-ASR.

\section{Speaker Embedding-aware Neural Diarization}

\label{sec:eand}
\begin{figure}[t!]
	\subfigure[SEND]{
		\includegraphics[width=8.5cm]{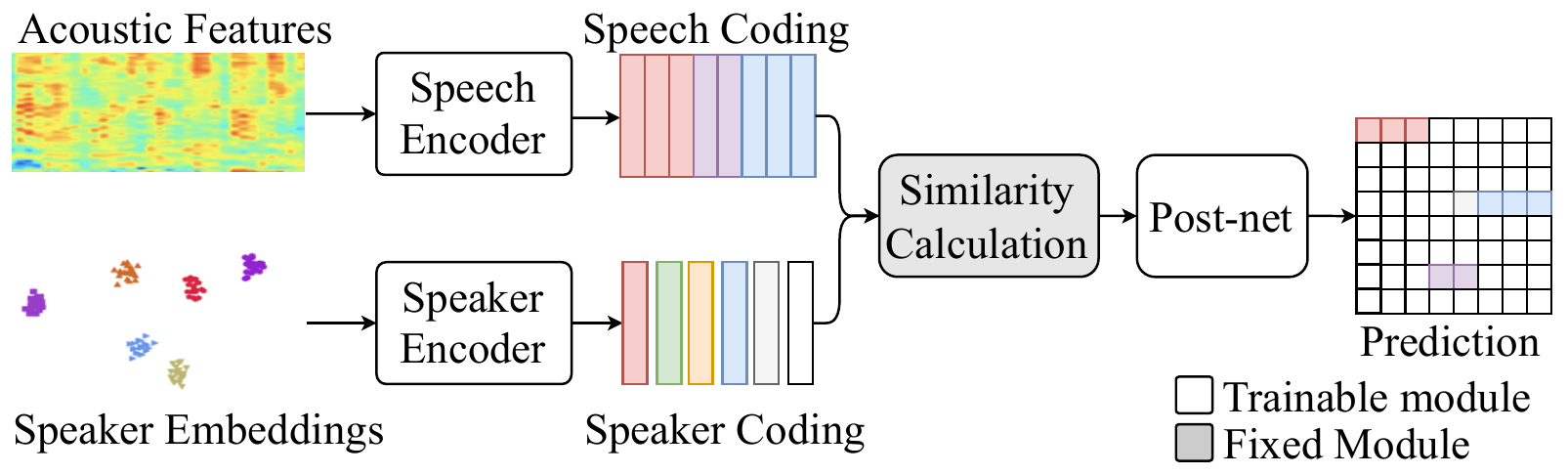}
		\label{fig:eand_a}
	}
	\subfigure[SEND-Ti]{
		\includegraphics[width=8.5cm]{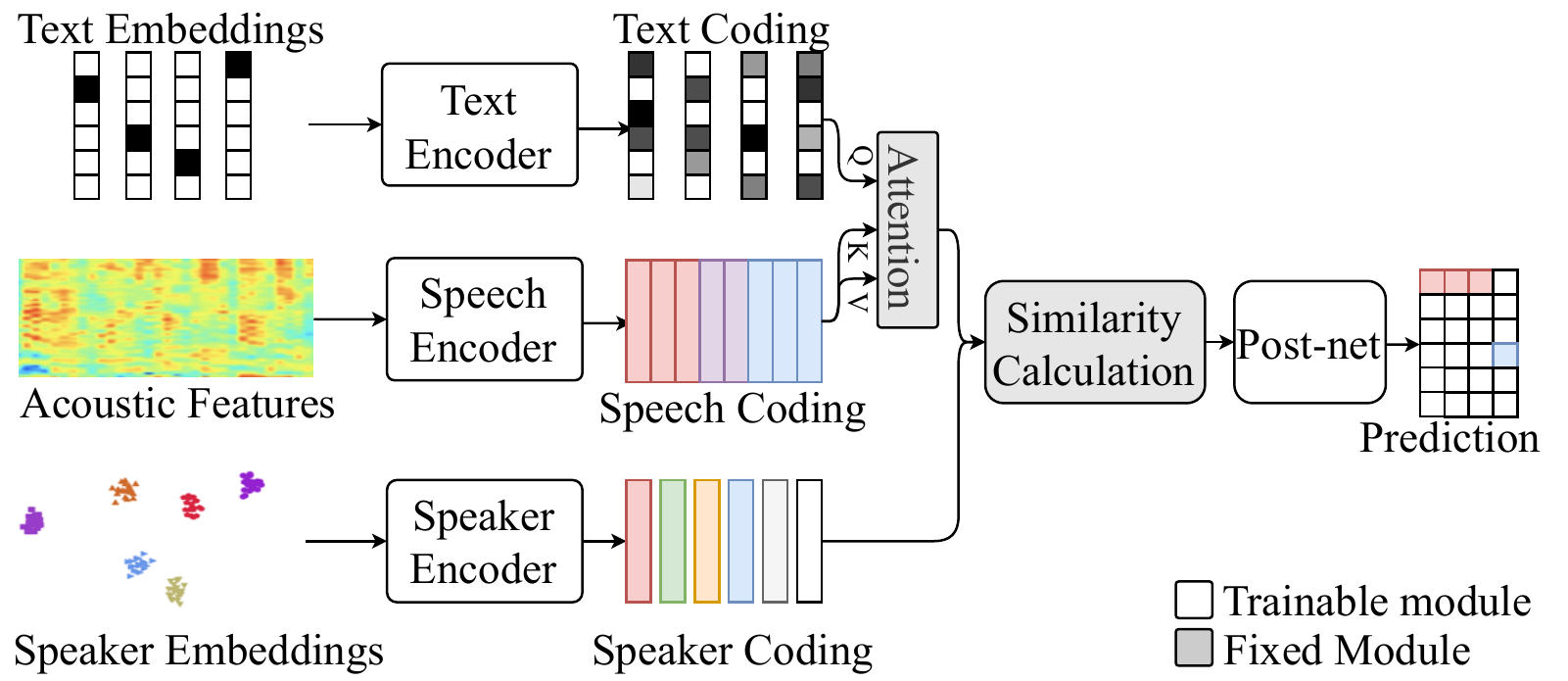}
		\label{fig:eand_b}
	}
	\caption{Semantic diagrams of (a) Speaker Embedding-aware Neural Diarization (SEND) and (b) SEND with textual information (SEND-Ti).}
	\label{fig:eand}
\end{figure}

As shown in Fig.\ref{fig:eand_a}, the proposed SEND consists of a speech encoder, a speaker encoder, a similarity calculation module and a post-processing network (post-net).
We further extend it with the textual information and obtain SEND-Ti, which comprises an extra text encoder and attention-based alignment module as shown in Fig.\ref{fig:eand_b}.

\subsection{SEND}
Given the acoustic features $X=\{\mathrm{x}_t|\mathrm{x}_t\in\mathbb{R}^F\}$ and speaker embeddings $V=\{\mathrm{v}_n|\mathrm{v}_n\in\mathbb{R}^D\}$, SEND aims at modeling the posterior 
probabilities of speaker activities at each time frame $P(Y|X,V)$, where $Y=\{\mathrm{y}_t|\mathrm{y}_t\in\{0,1\}^N\}$. 
However, $P(Y|X,V)$ can be intractable with a large number of frames and speakers. 
Therefore, it is factorized by using the conditional independence assumption:
\begin{equation}
\begin{split}
	P(Y|X,V) &= \prod_{t=1}^{T}{P(\mathrm{y}_t|\mathrm{y}_1,\dots,\mathrm{y}_{t-1},X,V)} \\
	         &\approx \prod_{t=1}^{T}{P(\mathrm{y}_t|X,V)}
\end{split}
\label{eq:post_fact}
\end{equation}
The frame-wise posterior probabilities $P(\mathrm{y}_t|X,V)$ are estimated through the speech encoder $\mathcal{H}$, speaker encoder $\mathcal{E}$, similarity calculation module and post-processing network $\mathcal{G}$:  
\begin{equation}
	P(\mathrm{y}_t|X,V) = \mathrm{SEND}(X,V;\mathcal{H},\mathcal{E},\mathcal{G})
\end{equation}

In SEND, the acoustic features $X$ are transformed into speech encodings $H=\{\mathrm{h}_t|\mathrm{h}_t\in\mathbb{R}^D\}$ 
through the speech encoder $\mathcal{H}$: $\mathrm{h}_t = \mathcal{H}(t,X)$.
Meanwhile, the speaker embeddings $V$ are transformed into speaker encodings $E=\{\mathrm{e}_n|\mathrm{e}_n\in\mathbb{R}^D\}$ through the speaker encoder $\mathrm{e}_n = \mathcal{E}(n,V)$. 
With the speech encodings $H$ and speaker encodings $E$, the pair-wise similarity matrix $A$ can be calculated:
\begin{equation}
	A_{t,n}=\mathbf{dot}(\mathrm{h}_t, \mathrm{e}_n), A_{t,n}\in [-\infty, \infty]
\end{equation}
where $\mathbf{dot}$ represents the dot product of two vectors. 
Since the dot product is not well-bounded, it can be sensitive to the outliers. 
We propose the activated dot product ($\sigma$-dot) by performing pre-activation before dot product:
\begin{equation}
\begin{split}
	A_{t,n} = \mathbf{dot}(\sigma(\mathrm{h}_t), \sigma(\mathrm{e}_n)) , A_{t,n}\in [-D, D]
	\label{eq:tanh_dot}
\end{split}
\end{equation}
where $\sigma$ is the hyperbolic tangent function. Another commonly used metric in speaker identification community is the cosine similarity between two vectors, 
which can be viewed as a normalized version of dot product:
\begin{equation}
\begin{split}
	A_{t,n} = \mathbf{dot}(\mathbf{Norm}(\mathrm{h}_t), \mathbf{Norm}(\mathrm{e}_n)), A_{t,n} \in [-1, 1]
\end{split}
\end{equation}
where Norm represents the l2-normalization.
Compared with the cosine similarity, $\sigma$-dot is aware with not only the angles but also the difference of length between two vectors. 

Given the pair-wise similarity matrix $A$, an approximation of posterior probabilities $P(Y|X,V)$ can be obtained by adopting the speaker-independent assumption:
\begin{equation}
\begin{split}
	P(Y|X,V) &\approx \prod_{t=1}^{T}{P(\mathrm{y}_t|X,V)} \approx \prod_{t=1}^{T}\prod_{n=1}^{N}{P(\mathrm{y}_{t,n}|X,V)} \\
	         &=\prod_{t=1}^{T}\prod_{n=1}^{N}{\frac{1}{1+\exp(-A_{t,n})}}
\end{split}
\end{equation}
The advantage of using this approximation is that the model can be trained and tested with a flexible and unknown speaker number $N$.
However, in real-life scenarios, speaker-independent assumption can be not satisfied. 
To model the potential relationship between speaker activities, a post-processing network $\mathcal{G}$ is involved to estimate the joint probability in each frame:
\begin{equation}
	\begin{split}
		P(Y|X,V) &\approx \prod_{t=1}^{T}{P(\mathrm{y}_t|X,V)} =\prod_{t=1}^{T}{\mathcal{G}(t,A)}
	\end{split}
\end{equation}

When the post-processing network is involved, the flexibility of SEND is also limited, since the model parameters always depend on the speaker number $N$.
This issue can be alleviated by setting $N$ to the maximum speaker number according to the prior knowledge \cite{he21c_interspeech}.
In both training and test phases, the speaker embedding matrix $V$ is augmented to match the hyper-parameter $N$ by randomly padding the zero vectors or the embeddings of negative speakers who do not talk anything in this segment.

\subsection{Power Set Encoding for Overlapping Speeches}
In previous methods, overlapping speech diarization is formulated as a multi-label problem, and the label $\mathrm{y}_{t,n}$ indicates whether speaker $n$ talks at frame $t$ or not.
There are two disadvantages in this formulation: i) the correlation between speakers is ignored, ii) a threshold need be set or tuned to transform the network outputs to the final diarization results.
To overcome these issues, we reformulate the overlapping speech diarization as a close-set single-label predication problem through the power set encoding (PSE).
Given $N$ speakers $\{1,2,\dots,N\}$, their power set $\mathcal{PS}(N)$ is:
\begin{equation}
\begin{split}
	\mathcal{PS}(N) &=\{S|S\subseteq \{1,2,\dots,N\}\} \\
	                &=\{\phi,\{1\},\{2\},\dots,\{1,2,n,\dots\}, \dots \}         
\end{split}
\end{equation}
where $\phi$ means the empty set. Subsequently, the elements in $\mathcal{PS}(N)$ are encoded according to the speaker order in matrix $V$:
\begin{equation}
	\mathcal{PSE}(S,N) = \sum_{n=1}^{N}{\delta(n,S)\cdot 2^{n-1}}
\end{equation}
where $\delta(n, S)$ is the Dirac function, which equals one if speaker $n$ belongs to set $S$ and zero otherwise.
By applying power set encoding on $N$ speakers, the number of labels can increase to $2^N$. 
Fortunately, in real-life scenarios, there are at most $K$ (e.g. two or three) speakers talking at the same time, and the number of valid labels $\mathcal{C}(K,N)$ is reduced to the summation of combinations:
\begin{equation}
	\mathcal{C}(K,N) = \sum_{k=0}^{K}{N \choose k} = \sum_{k=0}^{K}\frac{N!}{k!(N-k)!}
\end{equation}
Finally, the overlapping speech diarization task is reformulated as a single-label classification problem with $\mathcal{C}(K,N)$ labels:
\begin{equation}
	P(\mathrm{y}_t|X,V) = \mathcal{PSE}^{-1}(\mathbf{softmax}(\mathcal{G}(t, A))
\end{equation}
\subsection{SEND-Ti}
In this section, we extend the proposed method SEND to SEND-Ti by utilizing the textual information from the downstream automatic speech recognition systems or manual transcripts.
As shown in Fig.\ref{fig:eand_b}, word embeddings $Z$ are first encoded to textual encodings $U=\{\mathrm{u}_l|\mathrm{u}_l\in \mathbb{R}^D\}$ through an extra text encoder $\mathcal{U}$: $\mathrm{u}_l=\mathcal{U}(l, Z)$. Then, the acoustic information is aligned and aggregated into each textual encoding by the attention mechanism:
\begin{equation}
	\begin{split}
		\alpha_{l,t} &= \mathbf{dot}\left(W_q \mathrm{u}_l, W_k \mathrm{h}_t\right) \\
		a_{l,t} &= \frac{\exp(\alpha_{l,t})}{\sum_{i=1}^{T}\exp(\alpha_{l,i})} \\
		\mathrm{m}_l &= \sum_{t=1}^{T}{a_{l,t}W_v\mathrm{h}_t}
	\end{split}
\end{equation}
where $W_q$, $W_k$ and $W_v$ are trainable parameters. The pairwise similarities between each word and speaker are obtained in the same manners as equation (\ref{eq:tanh_dot}):
\begin{equation}
	A_{l,n} = \mathbf{dot}\left(\sigma(\mathrm{m}_l), \sigma(\mathrm{e}_n)\right) , A_{l,n}\in [-D, D]
\end{equation}
By passing the post-processing network $\mathcal{G}$, the final diarization results of each word are estimated:
\begin{equation}
	P(\mathrm{y}_l|X,V,Z) = \mathbf{softmax}(\mathcal{G}(t, A))
\end{equation}
Note that there is no overlapping problem in word-level diarization, therefore, the power set encoding is no longer used in SEND-Ti.
\section{Experimental Settings}
\label{sec:exp}

\subsection{Data}
\label{sec:illust}
We first perform the model and hyper-parameter selections on a small-scale simulation corpus.
Then a large-scale dataset is simulated to train and evaluate the model.
In addition, we also evaluate our method on an internal set, which is recorded in the real meeting scenarios with the overlap ratio of 31.81\%.
This real meeting set is also used in our previous study \cite{ZhengHWSFY21}.

To simulate the small-scale dataset, the AISHELL corpus \cite{BuDNWZ17} is employed, which consists of 170-hour speaker-homogeneous utterances recorded by 400 speakers.
We split them into 340 speakers for the training set, 40 speakers for the validation set and 20 speakers for the test set.

For the large-scale simulation dataset, an internal speaker verification corpus is used to simulate the multi-talker mixtures, which comprises about 30,000 speakers. 
As a result, the total duration of the large-scale dataset is about 6,000 hours. One percent of them is used as the validation set, and another percent is used as the test set.
Both small-scale and large-scale data sets are simulated in the similar manners as \cite{FujitaKHNW19}, but their maximum talker numbers in each mixture are set to two and four, respectively.



\subsection{Experimental Setups}
We extract 80-dimensional log-Mel-filterbank features with 25 ms frame length and 10 ms frame shift.
For the features in each frame, they are concatenated with three previous and succeeding frames, resulting in the dimension of 560. To deal with the long utterances, we sub-sample the features with a factor of six every seven frames.

We employ x-vector \cite{SnyderGSPK18} as the speaker embedding, which is extracted from the ``tdnn6\_dense'' layer of a time-delay neural network (TDNN). We train the TDNN with the additive margin loss function \cite{WangCLL18} and the original speaker-homogeneous utterances.
For each training sample of SEND, the embedding matrix $V$ consists of $n$ positive, $m$ $(0\leqslant m \leqslant N-n)$ negative and $N-m-n$ zero embeddings. While the positive embedding is extracted from the speaker who presents in the mixture, negative embeddings are extracted from the speakers who talk nothing in the mixture.
Note that the speaker embedding is extracted from a random enrollment utterance of 40 seconds, which can be slightly different for the same speaker in various samples. Preliminary experiments show that such data augmentation achieves better performance for short enrollments.

In the open meeting scenario, the enrollment utterances and their embeddings are hard to obtain. Therefore, we also evaluate the embeddings derived from the clustering results.
Specifically, we first obtain the clusters by using the VBx algorithm. Then, embeddings from the same cluster are averaged, and the clustering centers are employed as the input embeddings for SEND.

\subsection{Baseline Systems and Evaluation Metrics}
We employ a deep feed-forward sequential memory network (FSMN) as the speech encoder, which consists of eight FSMN blocks with 512 memory units.
A fully connected network (FCN) is adopted as the speaker encoder, which comprises three dense layers with 512 hidden units.
Inspired by the success of Transformer in text modeling \cite{VaswaniSPUJGKP17}, we employ a Transformer encoder to perform the text encoding in SEND-Ti, which comprises six self-attention blocks with 512 hidden units and eight heads.
Another FSMN is employed as the post-net to predict the multi-speaker labels or the power set encodings. The post-net comprises two FSMN blocks and $N$ memory units in each block.
The filter size of all FSMN blocks is 31. The hyper-parameter $N$ and $K$ are set to 16 and three, respectively.

We compare our methods with TSVAD \cite{MedennikovKPKKS20}, which is trained with the same acoustic features and speaker embeddings as SEND.
For fair comparison, the speaker detection and combining modules of TSVAD are replaced by FSMNs resulting the similar model size as SEND.
For the real meeting data, the VBx algorithm is also compared, which achieves superior performance on several popular datasets \cite{DiezBWRC19,landini2022bayesian}. 
All models are optimized 200,000 iterations by using the Adam optimizer with the learning rate of 1.0 and the batch size of 64.
The commonly-used diarization error rate (DER) is employed as the evaluation metric.
For SEND-Ti, we report the word-level DER (wDER) instead of DER, which is designed to measure the error caused in the lexical side \cite{abs-2101-09624}.

\section{Experimental Results and Analysis}
\label{sec:foot}

\subsection{Comparison of Different Similarity Metrics}
\begin{table}[t!]
	\centering
	\caption{The DERs (\%) of different similarity metrics on the simulation set.}
	\label{tab:comp_simi}
	\setlength{\tabcolsep}{7mm}{
	\begin{tabular}{lcc}
			\toprule
			Metrics & DER(Con.) & DER(Olp.) \\
			\midrule
		    cosine & 6.23 & 12.62 \\
			dot & \textbf{3.63} &  8.42 \\
			$\sigma$-dot & 4.23 & \textbf{7.87} \\
			\bottomrule
	\end{tabular}}
\end{table}
We first evaluate the impact of different similarity metrics with the same model architecture, and the results of DER are shown in Table \ref{tab:comp_simi}.
From the table, we can see that the dot similarity achieves the lowest DER on the concatenation set (Con.), in which the utterances from different speakers are concatenated without overlap.
However, when the utterances are overlapped, the proposed $\sigma$-dot similarity outperforms the cosine and dot similarity metrics with the relative improvements of 37.64\% and 7.24\%, respectively.
\subsection{The Effect of Post-net and Power Set Encoding}
\begin{table}[t!]
	\centering
	\caption{The DERs (\%) of different model architectures on the simulation set.}
	\label{tab:comp_pse}
	\setlength{\tabcolsep}{1.2mm}{
	\begin{tabular}{lcccc}
		\toprule
		Model & Post-net & $\mathcal{PSE}$ & DER(Con.) & DER(Olp.) \\
		\midrule
		TSVAD \cite{MedennikovKPKKS20} & - & - & 4.55 & 4.82 \\
		Exp 1 & None & $\times$ & 4.00 & 10.86 \\
		Exp 2 & FCN & $\times$ & 3.63 & 8.42 \\
		Exp 3 & LSTM & $\times$ & 4.32 & 6.27 \\
		Exp 4 & FSMN+FCN & $\times$ & 3.42 & 6.23 \\
		Exp 5 & FSMN+FCN & $\surd$ & \textbf{3.13} & \textbf{4.59} \\
		\bottomrule
	\end{tabular}}
\end{table}
We compare the model architectures of post-net and evaluate the effect of power set encoding in Table \ref{tab:comp_pse}.
By comparing Exp 1, 2 and 3, we can see that the diarization performance is further improved by concatenating a FSMN before the FCN. This indicates that the sequential modeling ability is crucial for the post-net. In speech processing community, another common choice for sequential modeling is the long short term memory (LSTM) \cite{HochreiterS97} recurrent neural network. Compared with LSTM, the ``FSMN+FCN'' model achieves better performance under both concatenation (Con.) and overlapping (Olp.) conditions with less parameters and lower computation complexity.
When the power set encoding is involved, the diarization performance is significantly improved.
Compared with the TSVAD, our method achieves 26.00\% and 4.77\% relative improvements on the concatenation and overlapping sets, respectively.
\subsection{Textual Information for Word-level Diarization}
\begin{table}[t!]
	\centering
	\caption{The world-level DER (\%) of different models on the simulation set.}
	\label{tab:send-ti}
	\setlength{\tabcolsep}{3.0mm}{
	\begin{tabular}{lcccc}
			\toprule
			Model & SC & Training Text & Grand & Recognition \\
			\midrule
			Exp 1 & $\times$ & Recognition & 3.12 & 3.28 \\
			Exp 2 & $\times$ & Grand & 2.97 & 3.19 \\
			Exp 3 & $\surd$ & Recognition & 1.82 & 2.08 \\
			Exp 4 & $\surd$ & Grand & \textbf{1.66} & \textbf{1.93} \\
			\bottomrule
	\end{tabular}}
\end{table}
The world-level DERs are shown in Table \ref{tab:send-ti}, where the ``SC'' column indicates whether the speaker change (SC) separator is contained in the text, and the ``Training Text'' column indicates that the model is trained with the grand-truth text from the human listener or the recognition text from an ASR system.
The ``Grand'' and ``Recognition'' columns indicate that the model is evaluated with the grand-truth and recognition text as inputs, respectively.
We train the ASR model with the serialized output training (SOT) technique \cite{KandaGWMY20}.

From Table \ref{tab:send-ti}, we can see that the world-level DER can be reduced significantly by involving the speaker change separator, no matter the model is trained with ground-truth or recognition text. Meanwhile, models trained with ground-truth text achieves better diarization performance than those trained with recognition text, which is because the ground-truth text can provide more accurate alignment between the textual and acoustic information. By comparing the results in column ``Grand'' and ``Recognition'', we find that the more accurate input text leads to less diarization errors. Compared the results in Table \ref{tab:comp_pse} and Table \ref{tab:send-ti}, we can see that the textual information can much benefit the speaker diarization system, which has been ignored in previous literature. We also try masking out the speech encodings from SEND-Ti, but obtain the DER of about 75\%. This indicates that it is difficult to recognize speakers with only textual information.
\subsection{Results on Real Meeting Data}
\begin{table}[t!]
	\centering
	\caption{The DER results (\%) of different models on the real meeting test set.}
	\label{tab:real_data}
	\setlength{\tabcolsep}{3mm}{
	\begin{tabular}{l|c|c|c|c}
		\toprule
		Embedding & \multicolumn{2}{|c|}{Oracle} & \multicolumn{2}{c}{Clustering}\\
		\midrule
		Model & Ignore & Full & Ignore & Full \\
		\midrule
		VBx \cite{landini2022bayesian} & - & - & \textbf{6.64} & 18.47 \\
		TSVAD \cite{MedennikovKPKKS20} & 17.15 & 21.99 & 15.47 & 21.80 \\
		SEND & 20.39 & 26.78 & 19.24 & 27.34 \\
		+Post-Net &18.51 & 22.64 & 17.19 & 22.08 \\
		\ \ +$\mathcal{PSE}$ & 11.09 & 16.06 & 10.54 & 16.44 \\
		\ \ \ \ +Adaptation & \textbf{7.46} & \textbf{10.76} & 8.08 & \textbf{12.17} \\
		\bottomrule
	\end{tabular}}
\end{table}
We also compare our method with VBx and TSVAD on internal real meeting data, which is a part of evaluation set in \cite{ZhengHWSFY21}.
For real data, we evaluate two speaker embeddings.
One is the oracle embedding which is extracted by using the rich transcription time marked (RTTM) file annotated by human listeners, the other is obtained from the clustering results of the VBx algorithm. While the oracle embeddings indicate the upper bound of our method, the clustering-based embeddings represent the diarization performance in real-world applications.
The DER results of different models are shown in Table \ref{tab:real_data}, where ``Ignore'' means the overlapping speeches are excluded from the DER calculation.
From the table, we can see that the proposed SEND method outperforms the TSVAD under all evaluation conditions by involving the post-net and power set encoding. Compared with VBx, our method achieves lower DER under the full test condition, where the errors of overlapping speeches are considered as well. By adapting the model with real meeting data, the diarization performance can be further improved. Note that the domain adaptation data is different from the test set. As a result, our method achieves 41.74\% and 34.11\% relative improvements for oracle and clustering-based embeddings, respectively. Interestingly, the ignore overlapping results of clustering-based embeddings are better than those of oracle embeddings, this is because the similar speakers are combined by VBx in some meetings.
\section{Conclusion}
\label{sec:conclusion}
In this paper, we reformulate the overlapping speech diarization as a single-label prediction problem, which is treated as a multi-label classification task in previous literature.
A novel framework, SEND, is proposed for flexible number of speakers. 
By involving the post-net and power set encoding, our method outperforms the TSVAD on both simulation and real meeting data sets.
Experimental results on the real meeting data show that, even only trained with simulation data, our method still achieves better performance than VBx algorithm, and the diarization errors can be further reduced through the domain adaptation.
Besides, we find that the textual information much improves the diarization performance for the downstream task, such as SA-ASR.
We will study the effect of textual information for the real meeting data in the future work.
\bibliographystyle{IEEEbib}
\bibliography{refs}

\end{document}